\documentclass[journal]{IEEEtran}
\usepackage{graphicx}
\usepackage{subfig}
\usepackage{mathtools, amsmath, amsthm, amssymb, amsfonts}
\usepackage{nicefrac}
\usepackage{tikz}
\usetikzlibrary{quantikz}
\usepackage{braket}
\usepackage{acronym} 
\usepackage{color}
\ifCLASSINFOpdf
\else
\fi
\usepackage{amsmath}
\usepackage{amssymb}
\usepackage{multirow}

\acrodef{Qubit}{Quantum bit}
\acrodef{RSA}{Rivest-Shamir-Adleman}
\acrodef{QFT}{Quantum Fourier Transform}
\acrodef{DFT}{Discrete Fourier Transform}
\acrodef{FFT}{Fast Fourier Transform}
\acrodef{MSB}{Most Significant Bit}
\acrodef{LSB}{Least Significant Bit}
\acrodef{TCS}{Technologies of Computing Systems}
\acrodef{MEEC}{Master on Electrical and Computer Engineering}
\acrodef{DSA}{Domain Specific Accelerator}
\acrodef{qubit}{quantum bit}
\acrodef{NISQ}{Noisy Intermediate Scale Quantum}

\begin{document}
%
\title{A Course-chapter on Quantum Computing for \\ Master's Students in Engineering}
%
%
%

\author{Leonel~Sousa,~\IEEEmembership{Senior Member, Computer Society Distinguished Contributor,~IEEE.}%
\thanks{Leonel Sousa is with the Department
of Electrical and Computer Engineering, Instituto Superior T\'ecnico,  University of Lisbon, and
at Instituto de Engenharia e de Sistemas de Computadores (INESC-ID), Portugal,
e-mail: \{leonel.sousa\}@inesc-is.pt.}}

\maketitle


\begin{abstract}
Quantum computing is a rapidly evolving field encompassing various disciplines such as physics, mathematics, computer engineering, and computer science. Teaching quantum computing in a concise and effective manner can be challenging, especially within the time constraints of a single course or a short period, even for graduate students. This challenge is particularly relevant in two-year MSc programs that include a thesis, which is a typical structure in higher education systems in the USA and Europe. In this paper, the author proposes an approach to teaching quantum computing and shares their experience of conducting a course-chapter on the subject within a two-week time frame. The experience reported in this paper is integrated into the "Technologies of Computing Systems" (TCS) course, with a total workload of 6 ECTS (in the context of the  European Credit Transfer and Accumulation System, one ECTS corresponds to 28 hours of work), conducted in one quarter, over seven weeks. The structure of the course-chapter is discussed, involving a series of lectures that were accompanied by lab classes and a lab project,  allowing students to receive guidance while also engaging in hands-on learning and independent study. The paper provides an overview of the quantum computing  topics covered, and their integration in the TSC course, and gives details about how these topics are studied in the different types of classes. It also discusses the evaluation procedure and presents the results obtained. It can be concluded that the inclusion of the quantum computing component not only significantly increased student interest in the course but also effectively bridged the gap between classical and quantum computing for engineering students within a short period of two weeks. This highlights the potential for successfully introducing and teaching quantum computing concepts in a compact format, in engineering programs, in particular in master programs in Electrical and Computer Engineering.   
\end{abstract}


%
\IEEEpeerreviewmaketitle

\section{Introduction}

\IEEEPARstart{W}ithin the last two decades, quantum technologies have progressed significantly, evolving  from quantum mechanics, and the study of the dynamics of particles at its most fundamental level, into a cross-disciplinary field of applied research~\cite{nielsen_2010}. 
Four different domains have been addressed with quantum technologies: {\it i}) communication, in which individual or entangled photons allow data transmission in a provable secure way; {\it ii}) simulation, where quantum systems are used to reproduce the behavior of other, less accessible, quantum systems; {\it iii})  sensing and metrology, which exploits the  high sensitivity of coherent quantum systems to external perturbations to measure physical quantities; and {\it iv}) computation, which employs quantum effects to set up the large computational state space and perform massively parallel computation. By using quantum superposition, $2^n$ inputs can be stored in $n$ quantum units of information, \acp{qubit}, simultaneously~\cite{Wong_2022}. 

In this last domain, quantum  computing takes a prominent place among the new computing paradigms that are currently being investigated. Moore’s prediction, which predicts that the number of transistors in digital electronics circuits  doubled every two years~\cite{Moore_2006} is currently no longer valid,  due to the physical characteristics of transistor technology, which reached dimensions at which quantum effects impact their functionality~\cite{Theis_2017}. Quantum technologies and quantum computing  have received significant attention not only from governments and public bodies, but also from private companies, some  large multinational companies, including Google, IBM, Intel, Microsoft, Toshiba, and other start-up companies. 

Quantum computing has already significant  domains of application on optimization, simulation, machine learning, and cryptography~\cite{Bayerstadler_2021}. To this end, quantum computing spans different complementary fields, such as physics, mathematics, computer engineering, and computer science. Students to be trained need a background in those areas. Recently,  researchers from across academia, government, industry, and national laboratories have proposed a modular quantum engineering course for the first year of undergraduate programs~\cite{Asfaw_2022}. Although it targets quantum-aware and quantum-proficient engineers at the bachelor’s level, there are undergraduate programs without specific courses, like the  proposed, on quantum information science and engineering.  Thus,  this paper presents  a  course-chapter on quantum computing for Engineering graduate students. This course is embedded in a more general course on \ac{TCS}, which is devoted to new paradigms of computation and emergent technologies for designing computing systems. It describes the experience of teaching this quantum computing course-chapter  as part of a 6~ECTS \ac{TCS} course in an Electrical and Computer Engineering (ECE) MSc program. The skills targeted with this course-chapter (2 ECTS)  are:

\begin{itemize}
    \item to identify quantum gates and analyze quantum circuits;
    \item to design quantum algorithms;
    \item to design and simulate circuits for quantum computation;
    \item to understand the limits of quantum technology and quantum systems.
\end{itemize}

The knowledge provided for supporting the skills  are: 

\begin{itemize}
    \item principles and math that support quantum computing;
    \item analysis of the operation of quantum gates and quantum circuits;
    \item derivation of quantum states and representation of data in the quantum domain;  
    \item tools for testing and designing circuits for quantum computation;
    \item designing quantum algorithms;
    \item understand quantum technology and quantum systems.
\end{itemize}

Lectures, interactive problem-solving, lab work, and a practical project are parts of a whole that make the learning process in this course very efficient. All these three components are tightly coupled, pushing forward the knowledge of graduate students and boosting their interest in quantum computing. 

The organization of this paper is as follows. Section~\ref{sec:organization} provides the context and the organization of the course, including how  the learning process is distributed  by the different types of classes, and grading, which allows us to assess the success of the proposal and of the course-chapter in practice. The subsequent Section~\ref{sec:topics}, presents the main topics of quantum computing covered by the course and how they are learned  and experienced in the classes. 
Associated also with this section is Appendix~\ref{appendix_math}   introducing students to  tensor algebra operations, which  are used to apply quantum gates and to compute states of a quantum system. 
Section~\ref{sec:technology} discusses the technology to implement quantum computing systems, based on the superconducting \ac{qubit} modality.  It corresponds to~\ac{NISQ} technology, in which non-error-corrected \acp{qubit} are used to implement quantum algorithms. Section~\ref{sec:labs} provides additional information about the labs and the practical project students implement, while Section~\ref{sec:assessment} provides information on how students are evaluated and the results of a query about the student's opinion and satisfaction with the course. Appendix~\ref{appendix_exam} provides one exam question on quantum computing of this year's edition of the \ac{TCS} course. Finally, Section~\ref{sec:Conclusions} draws the main conclusions and follows up on the impact of the  course on other activities of IST.

\section{Context and organization of the course}
\label{sec:organization}

Instituto Superior Técnico (IST), created in 1911, is the faculty of engineering of the Universidade de Lisboa (https://tecnico.ulisboa.pt/en/). 
The \ac{MEEC} is a  two years broad band MSc program offered by IST, with 120~ECTS, and seven  areas of specialization (https://fenix.tecnico.ulisboa.pt/cursos/meec21). 
It attracts around $200$ new students per year.

\ac{TCS} is an elective course of the MEEC with the main objective of providing students with the skills for understanding and designing computing systems based on emergent technologies. The course covers memory technologies and sub-systems, for example, resistive memories, and domain-specific accelerators, covering not only the technical issues but also the  economic impact~\cite{Thompson_2021}, and new paradigms of computation, such as processing in memory (PIM)~\cite{UPMEM_2019}~\cite{Yu_2021}, DNA-based computing, and quantum computing. A set of slideshows was prepared for introducing the key concepts, and students have a list of references  of technical papers and book chapters to support the learning process.  This paper only focuses on the chapter of the course related to quantum computing. 



At IST, the \ac{TCS} course runs in seven weeks (quarter). For the students, each of these seven weeks includes 4~hours of lectures and 3~hours of lab classes, with additional seventeen hours of autonomous work. \ac{TCS} is an elective  course with 6~ECTS, which corresponds to a total of 168~hours of student work. 
The last two weeks of the \ac{TCS} classes are fully devoted to quantum computing, which is an independent course-chapter. The total work hours of this chapter are 56~hours, 14~hours of contact with teachers,  and  42~hours of  autonomous work. Regarding the evaluation, $40\%$ of the final grade comes from a final written exam, and $60\%$ comes from two projects. Since its first edition in 2021/22, the \ac{TCS} course has been fully taught in English, including all material for lectures, labs, and projects are only available in English.
The last project, the one in quantum computing, occurs in weeks $6^{th}$ and $7^{th}$. The work in the first lab class of these two weeks is guided, introducing projects, materials, and simulators, while in the remaining classes, students have to do the project autonomously.
Details about the quantum computing syllabus are provided in the next sections.  

\section{Quantum computing: knowledge and skills}
\label{sec:topics}
This section presents the sequence of topics covered,  and the knowledge, and skills, acquired  in lectures and labs in a symbiotic way. From the start, students are introduced to software simulators to verify the operation of  the quantum gates, circuits, and algorithms, and  analyse  the quantum states. Two simulators are adopted in this course-chapter: the web-based Quirk simulator~\cite{quirk} and the Qiskit open-source software development kit~\cite{qiskit}. The former is a web-based simulator with a very user-friendly interface, that allows one to compose  quantum circuits by dragging and dropping components, but it is limited to  $16$ \acp{qubit}. The latter is able to handle circuits of up to $63$ \acp{qubit},  it is less intuitive to program and use, but has the advantage of having associated packages to interface with IBM Quantum systems. 
Given the nature and duration of this course-chapter, we advise students to use the Quirk simulator, but it is up to them which one. The study material is provided at the beginning of the course: Additionally to the set of slideshows,  exposing the basic concepts, references include the two main books~\cite{nielsen_2010} and~\cite{Wong_2022}. 

\subsection{Quantum bits and quantum states}
\label{sec:qubits}

The first concept introduced is the basic unit of information of a quantum computer. Starting from a classical bit (cbit), the two states of a cbit can be represented by '0' and '1' in the vector space: $'0' \equiv \left( \begin{array}{c} 1  \\ 0 \end{array} \right) \; \mbox{and} \;  '1'\equiv \left( \begin{array}{c} 0  \\ 1 \end{array} \right)$. Immediately,  this representation is extended so that the values in a  vector represent the probability of being in the states given by its location, index 0 and 1. Following this representation on the vector space, we can move from a cbit to a \ac{qubit} through~\eqref{eq:qubit}.

\begin{equation}
qubit \equiv \left( \begin{array}{c} p_0  \\ p_1 \end{array} \right) \;;\; \vert p_0 \vert^2+\vert p_1 \vert^2=1
\label{eq:qubit}
\end{equation}

Therefore, for complex probability amplitudes ($p_0, p_1 \in \mathbb{C}$), $\vert p_0 \vert^2$ represents  the probability density of a \ac{qubit} to be in the state '0', and simultaneously in '1' with probability density $\vert p_1 \vert^2$: the sum of the probability densities should be equal to one ($|p_0|^2+|p_1|^2=1$); for $p_0=1$ or $p_1=1$, we are in the classical domain, otherwise the \ac{qubit} is a superposition of both states. We can also consider the \ac{qubit} in ~\eqref{eq:qubit} as  a vector in a space defined by a set of orthogonal basis vectors~\eqref{eq:qubit_1}:
\begin{equation}
qubit \equiv p_0 \left( \begin{array}{c} 1  \\ 0 \end{array} \right) + p_1 \left( \begin{array}{c} 0  \\ 1 \end{array} \right) \equiv p_0 \vert 0 \rangle + p_1 \vert 1 \rangle \;,
\label{eq:qubit_1}
\end{equation}
with $\vert 0 \rangle$ and $\vert 1 \rangle$ being the Dirac's {\it ket} notation for states (vectors). In quantum mechanics, it is extensively used the Dirac {\it bra-ket} notation for computing the amplitude of going from an original state $\vert o \rangle$ to an ending state $\vert e \rangle$ through the inner product~\eqref{eq:bra-ket}, where $*$ means the complex conjugate and $^T$ the transpose of the vector.
\begin{equation}
    \langle o \vert e \rangle \equiv \vert o^* \rangle^T  \vert e \rangle
    \label{eq:bra-ket}
\end{equation} 

After learning how to represent a single \ac{qubit}, it is explained how to generalize the vector and Dirac representations for multiple \acp{qubit} by using the tensor product operation ($\otimes$). At this time, we introduce or revisit,  depending on the background of the students, basic tensor algebra, as presented in Appendix~\ref{appendix_math}. ~\eqref{eq:2-qubits} illustrates moving from $1$ to 2-\acp{qubit} $\vert pq\rangle$, noting that $\vert p_0 \vert^2+\vert p_1 \vert^2+\vert q_0 \vert^2+\vert q_1 \vert^2=1$. It is easy to generalize~\eqref{eq:2-qubits}  to n-\acp{qubit}. The normalization of the complex amplitudes means that the state of the system is a unit  in a $n$ dimensional complex vector space, which is Hilbert space.

\begin{eqnarray}
  \vert pq \rangle \equiv & \left( \begin{array}{c} p_0  \\ p_1  \end{array} \right) \left( \begin{array}{c} q_0  \\ q_1 \end{array} \right) = \left( \begin{array}{c} p_0 q_0 \\ p_0 q_1 \\ p_1 q_0 \\ p_1 q_1  \end{array} \right) \label{eq:2-qubits} \\
  {\mbox for} \;\; \vert 01 \rangle \equiv & \left( \begin{array}{c} 1  \\ 0  \end{array} \right) \left( \begin{array}{c} 0  \\ 1 \end{array} \right) = \left( \begin{array}{c} 0 \\ 1 \\ 0 \\ 0  \end{array} \right) \nonumber 
\end{eqnarray} 

When the concept of superposition is introduced, we postpone to the end of this section of the lectures, the description of two other important concepts: the {\it measurement effect} and the {\it quantum entanglement} between two \acp{qubit}. Regarding  measurement, it is highlighted that when the \ac{qubit} is measured it collapses to one of the basic states, which for a single \ac{qubit} is '0' or '1'. The measurement process irreversibly changes the state of a \ac{qubit} moving it away from superposition to a new state that is exactly the outcome of the measurement. Therefore, it implies that we cannot collect any additional information about the probabilities associated with an internal state  by repeating the measurement. 

Quantum entanglement is a complex phenomenon, difficult physics understanding. It occurs when a group of particles (\acp{qubit}) are generated or interact in such a way that the quantum state of a \ac{qubit} cannot be described independently of the state of others in the group, even when they are physically far away. Although no information is communicated,  the behavior of \acp{qubit} seems to be  somehow coordinated, such that measuring a \ac{qubit} also collapses the other in a correlated state. And this coordination happens even across vast stretches of space in an instantaneous way (considering the travel time of light between the entangled \acp{qubit}).  In this course-chapter, we don't go much further explaining the theory behind the entangled state, but when study quantum gates are studied, examples of \acp{qubit} in entangled states are provided.

\subsection{Graphical representation of states' space }
\begin{figure}[h]
    \centering
    \includegraphics[scale=0.7]{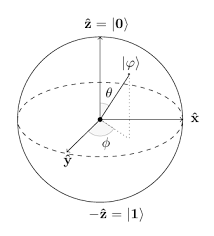}
    \caption{Bloch sphere}
    \label{fig:Bloch_sphere}
\end{figure}

It is useful to have insightful representations for the states of a \ac{qubit}, in particular through geometric and graphical means. For the case of \ac{qubit} states with $p_0$ and $p_1$ real numbers, the state of the \ac{qubit} can be represented as a unit circle in a bi-dimensional space, where $p_0$ and $p_1$ represent the $x$ and $y$ coordinates along the two orthogonal axes. Beyond  this particular set of states, students are taught about the possibility of representing the general case of $(p_0, p_1) \in \mathbb C^4$ in a 3-D space. An ingenious mapping,  natural but difficult to interpret, of a 4-D space, for representing two complex numbers, into a 3-D representation is achieved by introducing  the Bloch unitary sphere in Fig.~\ref{fig:Bloch_sphere}.  By applying  Euler's formulation of complex analysis to~\eqref{eq:qubit_1},  students obtain~\eqref{eq:Bloch_sphere_1}. By multiplying~\eqref{eq:Bloch_sphere_1} by $e^{-\phi_{p_0}}$, which does not change the probability of each state but only the phase, and by applying the Euler's formula leads to~\eqref{eq:Bloch_sphere_2}, and the representation in the unitary sphere with pair of angles ($\theta, \phi$) in  Fig.~\ref{fig:Bloch_sphere}.

\begin{equation}
\vert \varphi \rangle \equiv p_0 \vert 0 \rangle + p_1 \vert 1 \rangle \equiv R_0 e^{i\phi_{p_0}} \vert 0 \rangle + R_1 e^{i\phi_{p_1}} \vert 1 \rangle
    \label{eq:Bloch_sphere_1}
\end{equation}

\begin{eqnarray}
\vert \varphi \rangle &\equiv& \cos{(\theta/2)} \vert 0 \rangle+ \sin{(\theta/2)} e^{i\phi} \vert 1 \rangle    \label{eq:Bloch_sphere_2} \\
\phi & =&  \phi_{p_1}-\phi_{p_0} \;;\; \phi \in [0: 2\pi[ \nonumber\\
\theta & =&  2 \cos^{-1} R_0 =  2 \sin^{-1} R_1 \;;\; \theta \in [0:\pi] \nonumber
\end{eqnarray}

\label{sec:representation}

For all experiments conducted by the students, states are analysed in the Bloch sphere, for being familiar with this representation and identifying transitions between states. When a \ac{qubit} is in entanglement, its state cannot be individually represented. It is exceptionally considered to be in the center  of the sphere.

After these two sections of the course, students got the skills to represent the state of a quantum system through the state vectors of \acp{qubit}, and how to represent \acp{qubit} in the Bloch sphere. The next topic to teach is quantum gates, which define the transitions between quantum states.

\subsection{Quantum gates}
\label{sec:gates}

\begin{figure}
    \centering
    \includegraphics[scale=0.7]{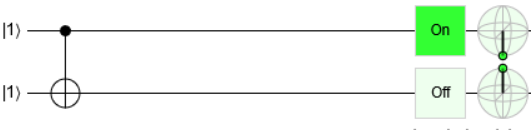}
    \caption{CNOT gate, with the input $\vert 11 \rangle$ (Quirk simulator~\cite{quirk})}
    \label{fig:CNOT_quirk}
\end{figure}

 Starting by explaining that like with classical digital circuits, gates in a serial datapath apply the operations sequentially, it is stated that quantum gates can be represented by matrices  and their operation by matrix multiplication over the vectors representing the quantum states. Thus, quantum gates operate over quantum states by applying linear algebra operators. This part of the course-chapter starts with students applying basic quantum gates, like the one in Fig.~\ref{fig:CNOT_quirk}. Students check the state's transition by hand, using tensor and linear algebra like in~\eqref{eq:CNOT}, and by simulation (Fig.~\ref{fig:CNOT_quirk}).

\begin{equation}
    {\mbox CNOT} \times \vert 11 \rangle  \Rightarrow  \left( \begin{array}{cccc}
        1 & 0 & 0 & 0 \\
        0 & 1 & 0 & 0 \\
        0 & 0 & 0 & 1 \\
        0 & 0 & 1 & 0 \end{array} \right) \times \left( \begin{array}{c}  0 \\ 0 \\ 0 \\ 1 \end{array} \right) =  \left( \begin{array}{c}  0 \\ 0 \\ 1 \\ 0 \end{array} \right)
        \label{eq:CNOT}
\end{equation}

\begin{figure}[h]
\begin{center}
\begin{quantikz}
\ket{1} &  \ctrl{1} & \qw &\qw \\ 
\ket{1} & \targ{} & \meter{meter} & \qw & '0'
\end{quantikz}
\end{center}
\caption{A meter (gate) outputs the value of the qubit  applying the measurement effect}
\label{fig:meter}
\end{figure}
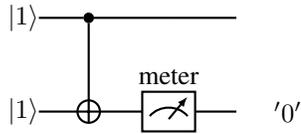

To measure the value of a qubit a  meter gate is applied, as illustrated in Fig.~\ref{fig:meter}. The measurement effect collapses the value of a qubit to one of the binary values '0' or '1'. Students realise  that beyond changing the value of the qubit, measurement is  not a reversible operation, whenever used it changes the value of a qubit and there is no way of moving back to the original value. It means that meters can not be used in a circuit for sensing intermediate values, only at the output to obtain the final result. 

The set of basic gates that students analyse and play with are represented in Fig.~\ref{fig:quantum_gates}.  All operations, implemented by gates, in quantum computing that do  not perform measurement are reversible and represented by unitary matrices. It means that for a quantum gate represented by matrix $\Psi$, $\langle \Psi^*\vert \Psi \rangle = {\bf I}$, with $\bf I$ being the identity matrix.  Like this, we make sure that for the  evolution of quantum systems, the sum of probabilities of all possible outcomes always equals the unit. 

\begin{figure}
    \centering
    \begin{tabular}{c c c} 
    Controlled Not (CNOT) & 
    \begin{tikzcd}
    \qw &  \ctrl{1} & \qw  \\ 
    \qw & \targ{} & \qw 
    \end{tikzcd}
&  $\left( \begin{array}{cccc} 
    1 & 0 & 0 & 0  \\ 
    0 & 1 & 0 & 0 \\ 
    0 & 0 & 0 & 1 \\ 
    0 & 0 & 1 & 0 \\ 
    \end{array} \right)$ 
    \vspace*{0.3cm} \\   
    Hadamard (H) & 
    \begin{tikzcd}
    \qw &  \gate[1]{H} & \qw 
    \end{tikzcd}
 &  
   $
    \left( \begin{array}{cc} 
    \frac{1}{\sqrt{2}} & \frac{1}{\sqrt{2}}  \\ 
    \frac{1}{\sqrt{2}} & -\frac{1}{\sqrt{2}} \\
    \end{array} \right)$
    \vspace*{0.3cm} \\
    Pauli-X (X) & 
    \begin{tikzcd}
    \qw &  \gate[1]{X} & \qw 
    \end{tikzcd}
 &  
   $ 
    \left( \begin{array}{cc} 
    0 & 1  \\ 
    1 & 0 \\
    \end{array} \right)$
    \vspace*{0.3cm} \\
    Pauli-Y (Y) & 
    \begin{tikzcd}
    \qw &  \gate[1]{Y} & \qw 
    \end{tikzcd}
 &  
   $ 
    \left( \begin{array}{cc} 
    0 & -i  \\ 
    i & 0 \\
    \end{array} \right)$
    \vspace*{0.3cm} \\
    Pauli-Z (Z) & 
    \begin{tikzcd}
    \qw &  \gate[1]{Z} & \qw 
    \end{tikzcd}
 &  
   $ 
    \left( \begin{array}{cc} 
    1 & 0  \\ 
    0 & -1 \\
    \end{array} \right) $
    \vspace*{0.3cm} \\
    Phase (S) & 
    \begin{tikzcd}
    \qw &  \gate[1]{S} & \qw 
    \end{tikzcd}
 &  
   $ 
    \left( \begin{array}{cc} 
    1 & 0  \\ 
    0 & i \\
    \end{array} \right)$ 
    \vspace*{0.3cm} \\
    $\pi/8$ (T) & 
    \begin{tikzcd}
    \qw &  \gate[1]{S} & \qw 
    \end{tikzcd}
 &  
   $ 
    \left( \begin{array}{cc} 
    1 & 0  \\ 
    0 & e^{\frac{\pi i}{4}} \\
    \end{array} \right)$ 
    \vspace*{0.3cm} \\
    Phase ($R_k$) & 
    \begin{tikzcd}
    \qw &  \gate[1]{R} & \qw 
    \end{tikzcd}
 &  
   $ 
     \left( \begin{array}{cc} 
    1 & 0  \\
    0 & e^{\frac{2\pi i}{2^k}} \\
    \end{array} \right) $
    \end{tabular}
    \caption{Basic quantum gates}
    \label{fig:quantum_gates}
\end{figure}


At this stage, students have to solve simple practical exercises in the lab classes. For example, to prove for one of the gates they choose  that it is unitary. And they play with the simulator to  check the transitions between states when different gates are applied in series.

\begin{figure}
    \centering
    \includegraphics[scale=0.7]{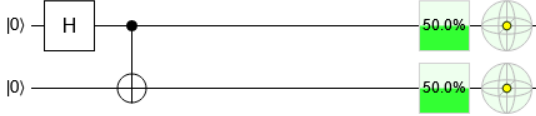}
    \caption{The control \ac{qubit} of a CNOT gate is placed in the superposition state: Bell state, simple example of entangled quantum states of two qubits}
    \label{fig:Bell_state}
\end{figure}

Students are exposed to practical exercises  to analyse the quantum circuit in Fig.~\ref{fig:Bell_state}. They compute the state's change produced by each of the two individual gates and compute the matrix corresponding to the whole circuit. The former solution is easier to compute, while the latter provides us a matrix that allows us to compute the behavior of the circuit for any input state. Jumping directly to this last approach, students reach~\eqref{eq:entanglement}. It is shown that having the state $\vert 00 \rangle$ at the input, the circuit produces  the output of the state given by~\eqref{eq:Bell_state_00}.

\begin{eqnarray}
H \otimes I & = \frac{1}{\sqrt{2}} \left( \begin{array}{cccc}
        1 & 0 & 1 & 0 \\
        0 & 1 & 0 & 1 \\
        1 & 0 & -1 & 0 \\
        0 & 1 & 0 & -1 \end{array} \right) \nonumber \\
 CNOT \times (H \otimes I) & = \frac{1}{\sqrt{2}} \left( \begin{array}{cccc}
        1 & 0 & 1 & 0 \\
        0 & 1 & 0 & 1 \\
        0 & 1 & 0 & -1 \\
        1 & 0 & -1 & 0 \end{array} \right)
 \label{eq:entanglement}   
\end{eqnarray}

\begin{equation}
 CNOT \times (H \otimes I) \times  \left( \begin{array}{c} 1 \\  0 \\  0 \\ 0 \end{array} \right)= \frac{1}{\sqrt{2}} \left( \begin{array}{c} 1 \\  0 \\  0 \\ 1 \end{array} \right)\label{eq:Bell_state_00}   
\end{equation}

It is quite interesting to observe that the state at the output in~\eqref{eq:Bell_state_00} not only represents superposition but also entangles the two quantum bits. According to to~\eqref{eq:Bell_state_00_entanglement}, it is not possible to separate  the behaviour of the two output quantum bits. It is shown through~\eqref{eq:Bell_state_00_entanglement} that the quantum state cannot be factored. Thus, students are introduced to the property of entanglement. The quantum bits are somehow coordinated, and measuring one \ac{qubit} collapses the other in a correlated way. The physical explanation for this phenomenon, which is controversial, is not detailed, but it is said that there is no communication, it can be probably applied to transmitting information very fast. At this stage, students are challenged to compute the states for the other three quantum input states and realize that they form Bell's states. These are specific quantum states of two \acp{qubit} that represent the simplest (and maximal) examples of quantum entanglement.

\begin{eqnarray}
 &  \frac{1}{\sqrt{2}} \left( \begin{array}{c} 1 \\  0 \\  0 \\ 1 \end{array} \right) \neq  \left( \begin{array}{c} a \\  b \end{array} \right) \otimes  \left( \begin{array}{c} c \\  d \end{array} \right), \forall (a,b,c,d) \in \mathbb{N} \nonumber  \\
 & (a=0 \vee d=0) \wedge  (b=0 \vee c=0) \Rightarrow \nonumber \\ 
 & (a.c=0) \vee (b.d=0) \;\; \mbox{proof of  the inequality} 
  \label{eq:Bell_state_00_entanglement}
\end{eqnarray}

At the end of this section of the course, students should have the tools and skills that allow them to analyse any quantum circuit with a defined number of \acp{qubit}. Moreover, they become familiar with the superposition and entanglement principles. 

\subsection{Quantum algorithms}
\label{sec:algorithms}

Within the time available in this course-chapter, a couple of seminal quantum algorithms are selected. It is shown how quantum computing allows to solve problems efficiently, and how they are subsequently implemented with quantum gates. The quantum selected algorithms, the Deutsch Oracle algorithm and the Shor algorithm are based on the \ac{QFT}. These  are also the algorithms  chosen  for students' projects in the lab classes.

The choice of  the Deutsch algorithm  has the purpose of showing  how quantum algorithms are developed and how parallelism is used to speed up the solution-finding process.
It is also important for showing that while parallel computations are performed, the information we can extract can be limited. The Deutsch Algorithm is a sub-instantiation for 1-\ac{qubit} of the more general quantum Deutsch–Jozsa algorithm, which provides an exponential speedup over a classic algorithm. For a n-\ac{qubit} problem, classically $2^{n-1}+1$ computations are required, but only $1$ computation using the general quantum algorithm. For the Deutsch algorithm $n = 1$, t thus reaches a speedup of two times.

The \ac{DFT} is an important discrete transform, used to perform Fourier analysis in many practical applications. It takes a finite sequence of equally-spaced input samples ($x$) into a same-length output sequence of equally-spaced samples input vector of samples ($y$).  The \ac{QFT} is the counterpart  of~\eqref{eq:DFT} for the quantum domain, and instead of complex numbers, it operates on quantum states. It is an important component of significant  algorithms, such as the Shor algorithm~\cite{Shor_1997} studied in this course-chapter.

\subsection{Deutsch Algorithm}

Students are exposed to the  concept of Oracle, observed as a black box but assuming that some physical system can be used to create it in the quantum computer (for example from simple quantum gates). The Deutsch oracle 
maps $f: \{0, 1\} \mapsto \{0, 1\}$, which means only one input bit, with  two possible values,  and one output bit, with also two values $f(0)$ and $f(1)$. The goal of the Deutsch algorithm is to find if the oracle is “balanced”, in this case ($f (0) \neq f (1)$), which corresponds to unary identity and negation, or “constant” ($f (0) = f (1)$). The quantum oracle $\mathbb{O}_f$ maps the input 2-qubit $\ket{x}\ket{y}$ into the output 2-\ac{qubit}s according to~\eqref{eq:Deutsch_Oracle}, where $\oplus$ represents the binary XOR operator. Students can check that it is easy to implement this oracle, it corresponds to  an identity gate, but that we do not know what is $f (x)$ in the oracle, we only know how it behaves.  With this oracle and the Deutsch algorithm, it is possible, with a single instantiation  by exploiting quantum parallelism, to identify if $f(x)$ is constant or if it is "balanced".  

\begin{equation}
\mathbb{O}_f \ket{x}\ket{y}=\ket{x}\ket{y \oplus f(x)}
    \label{eq:Deutsch_Oracle}
\end{equation}

To evaluate the function with different basis vectors at the same time, students are told to create a special superposition state at the input to the oracle. It is explained that by using Pauli-X ($X$) and Hadamard ($H$) gates (Fig.~\ref{fig:quantum_gates}),  superposition states  $\ket{+}\ket{-}$ are created leading to~\eqref{eq:Deutsch_superposition}.
\begin{eqnarray}
\ket{+}\ket{-}&=&\frac{1}{\sqrt{2}}(\ket{0}+\ket{1}) \otimes (\frac{1}{\sqrt{2}}(\ket{0}-\ket{1}) \nonumber \\
 &=&\frac{1}{2}(\ket{00}-\ket{01}+\ket{10}-\ket{11})
    \label{eq:Deutsch_superposition}
\end{eqnarray}

Following the proof of the algorithm as in~\cite{Wong_2022}, it is shown that by evaluating  the quantum Oracle (~\eqref{eq:Deutsch_Oracle}) the results in~\eqref{eq:Deutsch_result} are obtained for $f(x)$ constant or non-constant ($\overline{f(x)}$ is the negation of $f (x)$) .
\begin{eqnarray}
\mathbb{O}_f \ket{+}\ket{-}=\ket{+}\frac{1}{\sqrt{2}}(\ket{f(0)}-\ket{\overline{f(0)}}) \; \;\mbox{if}\;f(0)=f(1) \\
\mathbb{O}_f \ket{+}\ket{-}=\ket{-}\frac{1}{\sqrt{2}}(\ket{f(0)}-\ket{\overline{f(0)}}) \;\;\mbox{if} \;f(0)\neq f(1)
    \label{eq:Deutsch_result}
\end{eqnarray}

At the output of the Deutsch Oracle, it is easy to identify from the state of the first bit if $f(x)$ is constant or variable, the state will be $\ket{+}$ or $\ket{-}$, respectively. By applying a Hadamard gate ($H$), ($H\ket{+}=\ket{0}$; $H\ket{-}=\ket{1}$), the outcome of the Deutsch algorithm will be $\ket{0}$ for $f(x)$ constant, and $\ket{1}$ otherwise.

There are "lessons" that students can take  from the Deutsch algorithm, which are quite useful for quantum computing in general.
\begin{itemize}
    \item By taking a superposition of the basis states as an input to the Oracle, they are able to exploit quantum parallelism, and to achieve a speedup of $2$ even for this simple problem ($n=1$); with more \acp{qubit} (e.g. 40 \acp{qubit}), many computations can be performed at the same time (e.g. $2^{40}\approx 10^{12}$).
    \item While parallel computation is performed with the Deutsch algorithm, we only know if $f (x)$ is constant or balanced; it is the nature of many quantum computing algorithms,  which usually are  only able to extract a limited amount of information of the parallel processing, reached through superposition.
\end{itemize}

\subsection{Quantum Fourier Transform and Shor's Algorithm}
Starting from ~\eqref{eq:DFT}, in the quantum domain, complex numbers are replaced by quantum states.  An arbitrary quantum state $\sum\limits_{i=0}^{N-1}x_i\ket{i}$ is mapped through the \ac{QFT} to the state $\sum\limits_{i=0}^{N-1}y_i\ket{i}$ according to~\eqref{eq:QFT}.

\begin{eqnarray}
    y_k &=& \frac{1}{\sqrt{N}}\sum\limits_{j=0}^{N-1}x_je^{-jk \frac{2\pi i}{N}} \label{eq:DFT} \\
    QFT(\ket{x}) &=& \frac{1}{\sqrt{N}}\sum\limits_{y=0}^{N-1}e^{-xy \frac{2\pi i}{N}}\ket{y} \label{eq:QFT}
\end{eqnarray}
It is important to note that the quantum states  in~\eqref{eq:QFT} should be represented in binary notation, with each \ac{qubit} corresponding to a single bit.

As it is usually done for the \ac{DFT},~\eqref{eq:QFT} can be presented in a  matrix  form. By considering ${\bf W}= e^{i\frac 2\pi}{N}$ it leads to~\eqref{eq:QFT_W}.

\begin{equation}
QFT(\ket{x}) =  \left( \begin{array}{cccc}
        W^{0} & W^{0} & \cdots & W^{0} \\
        W^{0} & W^{1} & \cdots & W^{N-1} \\
        \vdots &  \vdots &  \ddots &  \vdots \\
        W^{0} & W^{N-1} & \cdots & W^{(N-1).(N-1)} \end{array}\right)\ket{y}
\label{eq:QFT_W}
\end{equation}

By deriving in detail the calculations for a 2-\ac{qubit} system, it is shown that, similarly to  the \ac{FFT}, it is possible to achieve a faster version, the \ac{QFT} with  iterative computation~\cite{nielsen_2010}. The general equation for iteratively calculating \ac{QFT} for $N=2^n$ is~\eqref{eq:FQFT}.

\begin{eqnarray}
    & QFT(\ket{x_1 \cdots x_n}) = \frac{1}{\sqrt{N}}\bigg[
    (\ket{0} + e^{2\pi i 0.x_n}\ket{1})\otimes \nonumber \\
    & (\ket{0} + e^{2\pi i0.x_{n-1}x_{n}}\ket{1})
    \cdots\otimes
    (\ket{0} + e^{2\pi i0.x_1...x_{n}}\ket{1})\bigg].
    \label{eq:FQFT}
\end{eqnarray}

By considering the  controlled-R quantum gate that applies a relative phase change to the state $\ket{1}$, as presented in Fig.~\ref{fig:quantum_gates}, a quantum  circuit for implementing the algorithm of the \ac{QFT} can be found by the students in~\cite{nielsen_2010}.




The \ac{QFT} Quantum Fourier Transform (QFT) is crucial for several quantum algorithms, in particular for the Shor’s algorithm~\cite{Shor_1997} analysed by the students. Shor’s algorithm is used for finding the prime factors of an integer ($N=P \times Q$) and thus could be used to break public-key cryptography schemes, such as the RSA.  Shor’s Algorithm factorizes a number ($N$) by guessing one of its factors and improving that guess in case it is not. Shor’s Algorithm starts by taking a random guess $P'$, and checking if it shares a factor with the large number $N$ using Euclid’s algorithm. If that is the case, the other factor can be simply obtained as $P=P' \;, \; Q=N/P$. However, since $N$ is the product of two very large ”random” prime numbers,  finding a factor is extremely unlikely and so, the algorithm will transform $P'$ into a better guess. This new guess comes from the application of Euler’s theorem to the relation between $P'$ and $N$: if $P'$ is not a factor of $N$, they are coprime numbers and through Euler’s theorem can be written as~\eqref{eq:Shor_algorithm}.

\begin{equation}
\mathop{P,N}_{(coprimes)} \rightarrow  P'^g=mN+1 = (P'^{\frac{g}{2}}+1)(P'^{\frac{g}{2}}-1)=m \times N    
    \label{eq:Shor_algorithm}    
\end{equation}

The improved guess is of the form $(P'^{\frac{g}{2}} \pm 1)$, only the value of $p$ has to be determined. For the equality $P'^g=mN+1$ in~\eqref{eq:Shor_algorithm},  we can consider the function  $f(x) = P'^x \mod{N}$, that  is periodic, with period g, being the period  given by the smallest integer that respects  $P'^gp \mod N = 1$. Finding $p$ is not an easy task for a large $N$, this is what assures the  security of modern cryptography. The difficulty comes from the fact that finding the periodicity of $f(x) = P'^x \mod N$ requires separately computing each value of $f(x)$. However, in a quantum system, this issue can be overcome thanks to quantum superposition, which allows the computation of all the images of $f(x) = P'^x \mod N$ at the same time. Given that f(x) is periodic, the resulting state is composed of equal values separated by the period. When the \ac{QFT} is applied, the results  state is composed of the \ac{QFT} of every image of $f(x)$, which adds together in a state that contains the frequency at which each value repeats $\ket{1/p}$.



By the end of this section of the course, students have the skills to analyse quantum algorithms and derive quantum circuits to implement them. Some of them  start thinking about developing algorithms for known problems, but the majority need to do more work to reach that level.

\section{Quantum computing technology and systems}
\label{sec:technology}

It is not possible in such a short period of time to study the physics of the quantum phenomenon and, in detail, the technology behind quantum computers. Not addressing different types  of technologies, we have decided to focus the attention on superconducting technology, and quantum computers based on this technology, pointing out the paper~\cite{Kjaergaard_2020} for students to read on this topic. 

The superconducting \ac{qubit} has been used on the~\ac{NISQ} technology, in which non-error-corrected \acp{qubit} are used
to implement quantum  algorithms. Superconducting \ac{qubit}-based quantum computers are currently dominant, particularly in industry. Recent advances on two-\ac{qubit} gates implementation, as well as operations on logical \acp{qubit} in extensible
superconducting \ac{qubit}-based quantum computers, shows this technology  also holds promise for the longer-term goal of building larger-scale error-corrected quantum
computers.~\cite{Kjaergaard_2020}. This course  discusses the recent experimental
advances in \ac{qubit} hardware, gate implementations, readout
capabilities, early~\ac{NISQ} algorithm implementations, and quantum error
correction using superconducting \acp{qubit}. 

This course-chapter introduces the basic devices for designing  superconducting \ac{qubit} systems and discusses the concept of coherence, the status when  the system's behavior can be explained by quantum mechanics.

\begin{figure}
    \centering
    \includegraphics[scale=0.85]{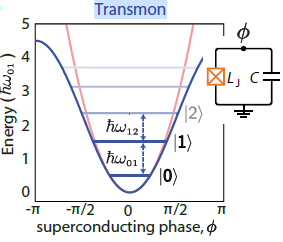}
    \caption{Transmon \ac{qubit}, with a Josephson junction~\cite{Kjaergaard_2020}}
    \label{fig:Transmon}
\end{figure}

Contrary to the energy spectrum of a $LC$ harmonic oscillator, the energy spectrum of the transmon \ac{qubit} in Fig.~\ref{fig:Transmon} with a  non-linear Josephson junction produces non-equidistant energy levels. The Josephson junction consists of two superconducting electrodes separated by a thin insulating barrier. It allows 
for the coherent tunneling of Cooper pairs, which results in a non-linear inductor.
Typically, the two lowest levels are used to define a qubit, with $\ket{0}$ corresponding to the ground state and $\ket{1}$  to the excited state. Superconducting circuits have evolved in the last years from two different  types of qubits,  the one based on electric charge and the one based on magnetic flux. The lifetime of a transmon \ac{qubit} is on the order of magnitude of hundred microseconds (Fig~2 of~\cite{Kjaergaard_2020}).

The manufacturing of superconducting circuits is  a multi-step additive and subtractive fabrication process, which involves  lithographic patterning, metal deposition, etching, and controlled oxidation of films of a superconductor, such as aluminum or niobium. Circuits are fabricated on silicon or sapphire substrates, leveraging techniques compatible with silicon CMOS manufacturing. Devices are placed inside a copper or aluminum package that provides an engineered electromagnetic environment thermally anchored to the $\approx 10$mK (less than $-270^o$C) stage of a dilution refrigerator. Additionally  to the \acp{qubit}, the superconducting circuits comprise resonators and bias lines. 

Developing~\ac{NISQ} algorithms will rely on access to quantum computers
of increasing complexity, some of them with cloud-based access to superconducting
quantum systems that enable algorithm designers and students to test ideas. 
IBM pioneered with a 5-transmon \ac{qubit} device in 2016, and since then IBM systems have improved. Nowadays, IBM gives access,  for example, to  D-Wave (see https://cloud.dwavesys.com/). Microsoft (https://azure.
microsoft.com/en-us/services/quantum/), Amazon (https://aws.amazon.com/braket/),
and Google have also announced plans to enable cloud access to their quantum systems. Google
recently developed a 53-\ac{qubit} processor, named Sycamore, and reports the first demonstration of quantum computational supremacy, by  solving a problem using a quantum
computer significantly faster than the best-known algorithm on a classical computer.
Sycamore comprises 53 individually controllable transmon-type \acp{qubit} and 86 couplers used to turn on/off nearest-neighbor 2-\ac{qubit} interactions.

\section{Labs and projects}
\label{sec:labs}
\begin{figure*}
	\center
	\subfloat[]{\includegraphics[width=0.056\textwidth]{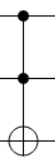}}\hspace{0.1cm}
	\subfloat[]{\includegraphics[width=0.85\textwidth]{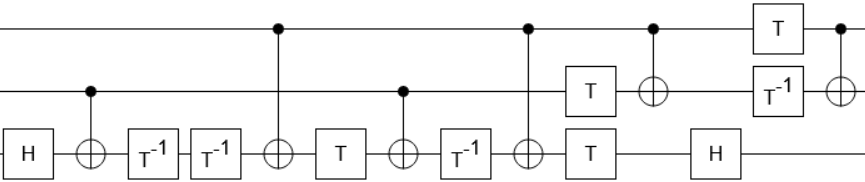}}
	\caption{TOFFOLI gates are decomposed into six CNOT gates and several one-\ac{qubit} gates~\cite{Shende_2009}}
	\label{fig:CCNOT}
\end{figure*}

A Lab class lasts for one and a half hours, and students have two of these classes per week, which leads to a  total of four classes and six hours. Students organize themselves in groups of three. There is a Teaching Assistant (TA) in any of these classes, but only in the first of these classes, the TA plays an active role. In the first class, Quirk and Qiskit simulators are introduced, the advantages and disadvantages of each one are discussed, and a simple demo of using each one to describe simple quantum gates and circuits is provided. Students have until the second class to choose and announce which simulator they are going to use. Students have to analyse, outside classes, the  operation of a more complex quantum gate by using the simulator. An example of such a gate is the {\it Toffoli}, also known as the CCNOT gate, represented in Fig.~\ref{fig:CCNOT}. The {\it Toffoli} gate is decomposed into six CNOT gates and several one-\ac{qubit} gates, which has been  demonstrated to be a  CNOT-based optimal implementation~\cite{Shende_2009}--\ac{qubit} gate $T^{-1}$ is the complex conjugate of $T$, which observing Fig.~\ref{fig:quantum_gates} corresponds to a rotation of $-45^o$ when the input \ac{qubit} assumes the state $\vert 0 \rangle$. The simulation of this more complex quantum gate is the  opportunity students have to play with the simulator, analyse the state change induced by basic quantum gates, and understand the information provided by the Bloch sphere. In all other lab classes, students are developing a project in an autonomous way.

A couple of quantum algorithms  are suggested to students, they choose the one they found  more interesting to develop, derive the circuit and simulate its operation. The suggested work have always two stages, a simple first stage that is close to what they easily find in the study book, and a second one that challenges students to go further in the topic. In the first edition of the course, the suggested quantum algorithms are:
\begin{itemize}
    \item a binary adder in the quantum domain, considering at the input only the two states $\ket{0}$ and $\ket{1}$; in this first stage of the work, students only need to use CNOT and {\it Toffoli} gates~\cite{sousa_2021}; in this first stage, students have also to extend the quantum circuit to implement an adder with two-input \acp{qubit}; the challenge for the second stage of the work is to design adding circuits but taking advantage of the superposition, for example, based on the QFT.   
    \item the Deutsh Oracle and the Deutch algorithm to identify the class of unitary operators in the circuit ($f(x)$ constant or balanced); for the second stage of the work, students are challenged to extend the algorithm from $1$ to $2$ bits, which means to extend the work to the Deutsch–Jozsa algorithm.  
\end{itemize}

Some of the students have decided to run the simple quantum algorithms they have developed on cloud-based superconducting quantum systems, such as the one made available by IBM. In the week after the last lab class, students are required to deliver a report, with a maximum of 4 pages, formatted according to the two-column IEEE manuscript templates for conference proceedings. The structure is the one of a paper, with the following sections. 
\begin{itemize}
    \item Introduction: explaining  the aim of the work in the context of quantum computing.
    \item Methodology: analysing and discussing the simulated circuits.
    \item Results: describing the attained results and drawing the main conclusions.
    \item Conclusions: summarizing the work, including the approach and results.
\end{itemize}

\section{Assessment and feedback from students}
\label{sec:assessment}

For the course, the assessment, in addition to a final exam, requires a demonstration of the operation of the designed quantum circuits and an individual discussion with each group of the different aspects of the work and the options taken. This demonstration is made in the last laboratory session, and  the  report has to be delivered one week after. 

The experience of the two  years offering this course-chapter has shown that some, but few, students show difficulties to face this completely new approach to computing. These difficulties arise mainly when students have to integrate knowledge of this course-chapter to extend quantum algorithms for solving more complex problems.  This is a component of the course-chapter that must be reinforced if one more week would be available. The group work and the practical exercises and tools have  been revealed to be essential to overcome difficulties. The \ac{TCS} course, like the majority of the courses in \ac{MEEC}, is elective and offered in the first of the two years of the master's program. 

IST runs a system that monitors the quality of the courses on a yearly  basis,  called Quality of the Curricular Units (QUC). The results of the QUC are based on the final grades, on the results of questionnaires anonymously answered by students about the organization of the courses, the material available to study and to perform experimental work, and also about the availability of the professor to support students and follow their work. Moreover, since it is an elective course, the attractiveness  also reveals the real interest that the course rouses in students.

In the first edition, the course attracted 58 students, while in the second edition, this number rises to 72 (a raise of $24\%$) out of a total of 200 students that can choose it. The percentage of students approved, considering the two editions, is around $90\%$ and the average grade is $15.5$ out of $20$.  In fact, in this second edition, it became the most popular course in the two specialization areas  offered, {\it Computing Systems} and {\it Cyber-physical Systems}. In the second edition, also foreign students have enrolled also in the course, for example, students from NTNU in Norway, {\it Université de Grenoble} in France, and {\it  Hochschule München University of Applied Sciences} in Germany. According to the students, the course-chapter on quantum computing is the main responsible for these results. Moreover, students have given the maximum grade to the course, 9 out of 9. The course is awarded with the seal of  Excellent by IST.

\section{Conclusions and follow up}
\label{sec:Conclusions}

The course-chapter on quantum computing presented in this paper  was designed  for engineering students and tested with quite a success over two consecutive academic years in an MSc program on Electrical and Computer Engineering. It is an intensive course, with total work hours of $56$~hours, $14$~hours of contact with teachers,  and  $42$~hours of  autonomous work,  in two weeks. It took a while to design a course-chapter that focuses on the main aspects of quantum computing and, at the same time, gives the opportunity to students to practice this non-conventional and not mature computer technology. Abstracting the computation on most of the course, and only at the very end tackling the technological aspects is a good approach. It allows students to practice quantum circuits and algorithms, during the two weeks, without facing  the additional difficulties of quantum technology, including quantum decoherence. Moreover, interleaving lectures with lab classes and practical projects, in a very synchronous way, have motivated students for studying and learning the main topics and to surpass the novelty that this course-chapter carries in terms of quantum computing fundamentals and technology. Students evaluated the quality of this elective course as excellent and the number of students that choose the course increased by  $24\%$ in the second edition of the course. Students from foreign universities were enrolled in the MSc program under the European Erasmus mobility program. Moreover, some students that have attended the \ac{TCS} course later on enrolled in the Minor in Quantum Science and Technologies at IST, which offers three courses  in the field of Quantum Technologies, to deepen their knowledge of quantum technology. Some of these  students chose to make  MSc theses in quantum computing and quantum technologies. An example was an MSc thesis that analysed in depth Shor's algorithm and implemented it by using  Qiskit, while some of the quantum circuits were tested on the IBM quantum computer~\cite{Valentim_2021}. This course arouse the interest in quantum computing at IST, so students that attended the course  have organized  an in-person  discussion panel on quantum computing in the 2022 edition of the  {\it IST Electrical and Computer Engineering Students Conference} (https://deec.tecnico.ulisboa.pt/noticias/comunidade/jeec-2022). I have moderated this panel, with the participation of well-known  international researchers in this domain, Koen Bertels, Professor at the University of Ghent and  founder of "QBee.eu", and Yasser Omar, Physicist expert on  quantum technologies and Professor at IST. Looking ahead, for computer science students it would be good to reinforce the algorithmic component, while for electrical engineering students, it is important to deepen the technological component. For both, it is important to understand the restrictions imposed by~\ac{NISQ}technology, in which non-error-corrected qubits are used to implement quantum algorithms. Moreover, to be aware of the longer-term effort of building larger-scale error-corrected quantum
computers. 

\appendices

\section{Additional Basic Math Topics}
\label{appendix_math}

This course assumes students have a background in simple linear algebra operations, such as matrix-vector and matrix-matrix  addition and multiplication. We provide additional information on tensor algebra, in particular product that is important for quantum computing and may not be taught in the basic courses of engineering BSc programs.   

 Given two vectors $\bf v$ and $\bf w$, the tensor product $\bf y$ is calculated as the external product $\bf y = \bf v \otimes \bf w = \bf v \bf w^\top$, where $\bf{w}^\top$  means the vector $\bf w$ transposed (and if the entries of $\bf w$ are complex numbers, then each 
 entry is also replaced by its complex conjugate). So, for a $\bf v$ in the space $\mathbb{R}^3$ and $\bf w$ in the space $\mathbb{R}^2$, the tensor product $\bf y$ in the $\mathbb{R}^3 \otimes \mathbb{R}^2$ is computed by~\eqref{eq:vector_tensor_product1}.

 \begin{equation}
     \bf y= \left( \begin{array}{c}
          v_1\\
          v_2 \\
          v_3
    \end{array} \right)_{3 \times 1} \otimes \left( \begin{array}{c}
          w_1\\
          w_2     \end{array} \right)_{2 \times 1} = \left( \begin{array}{cc}
               v_1 w_1 & v_1 w_2 \\
               v_2 w_1 & v_2 w_2 \\
               v_3 w_1 & v_3 w_2 
          \end{array} \right)_{3 \times 2}
\label{eq:vector_tensor_product1}
 \end{equation}

Any $m\times n$ matrix can be reshaped into a $nm\times 1$ column vector and vice versa. So, we exploited the fact that $\mathbb{R}^3 \otimes \mathbb{R}^2$ is isomorphic to $\mathbb{R}^6$ through~\eqref{eq:vector_tensor_product2}. 
 
 \begin{equation}
     \bf y= \left( \begin{array}{c}
          v_1\\
          v_2 \\
          v_3
    \end{array} \right)_{3 \times 1}  \otimes \left( \begin{array}{c}
          w_1\\
          w_2     \end{array} \right)_{2 \times 1} = \left( \begin{array}{c}
               v_1 w_1 \\
               v_1 w_2 \\
               v_2 w_1 \\ 
               v_2 w_2 \\
               v_3 w_1 \\ 
               v_3 w_2 
          \end{array} \right)_{6 \times 1}
\label{eq:vector_tensor_product2}
 \end{equation}

The dot product  of the  $n \times m$ matrix $\bf V$  by the $n \times m$ matrix $\bf W$ can be obtained by the dot product of vectors that compose the matrix. It leads to the matrix $nm \times nm$ matrix $\bf Y$ in~\eqref{eq:matrix_tensor_product2}.

\begin{eqnarray}
     & \bf Y  = \bf V \otimes \bf W = \left( \begin{array}{cc}
          v_{11} & v_{12} \\
          v_{21} & v_{22}
    \end{array} \right)_{2 \times 2} \otimes \left( \begin{array}{cc}
          w_{11} & w_{12} \\
          w_{21} & w_{22}
\end{array} \right)_{2 \times 2}  \nonumber\\
     & \bf Y  = \left( \begin{array}{cc}
               v_{11} \left( \begin{array}{cc}
          w_{11} & w_{12} \\
          w_{21} & w_{22}
\end{array} \right) & v_{12} \left( \begin{array}{cc}
          w_{11} & w_{12} \\
          w_{21} & w_{22}
\end{array} \right) \\
           v_{21} \left( \begin{array}{cc}
          w_{11} & w_{12} \\
          w_{21} & w_{22}
\end{array} \right) & v_{22} \left( \begin{array}{cc}
          w_{11} & w_{12} \\
          w_{21} & w_{22} \end{array} \right) \end{array} \right)_{4 \times 4}
\label{eq:matrix_tensor_product2}
 \end{eqnarray}

\vspace*{0.5cm}\section{Typical Question of an Exam}
\label{appendix_exam}

This section provides an example of the question of the \ac{TCS} exam on quantum computing. The number of points assigned to this question is $6.5$, which means one third of the $20$ total points, approximately the weight of this component of the course.

\begin{center}
\begin{table}[h]
\hrule
\vspace*{0.1cm}
\begin{tabular}{p{6.5cm}p{4cm}}
\textbf{Technologies of Computing Systems} & \multirow{3}{0pt}{\includegraphics[scale=0.08]{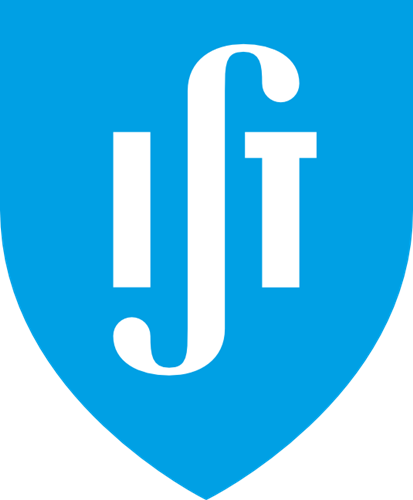}}  \\
   \textbf{Exam}, 26 January 2023 & \\
   &
\end{tabular}
\vspace*{0.2cm}
\hrule
\end{table}
\end{center}
\vspace*{-1cm}
\noindent 1) For the quantum circuit based on Hadamard ($\bf{H}$) and CNOT quantum gates
\begin{center}
	\includegraphics[scale=0.5]{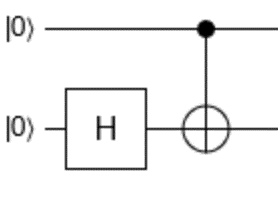}\\
\end{center}

\noindent [2.5 points] a) compute the value of the output \acp{qubit} when the input \acp{qubit} assume both the value $\vert 0 \rangle$.  

\vspace*{0.2cm}
$\bf{H}  \equiv \frac{1}{\sqrt{2}} \begin{pmatrix}
1 & 1 \\
1 & -1
\end{pmatrix}
$; $
CNOT  \equiv  \begin{pmatrix}
1 & 0  & 0 & 0 \\
0 & 1  & 0 & 0 \\
0 & 0  & 0 & 1 \\
0 & 0  & 1 & 0
\end{pmatrix}
$.

\vspace*{0.3cm}
\noindent [1.5 points] b) Is the circuit an entangled system? in an entanglement state? Justify the answer and, if it is not the case, give an example of a circuit with quantum entanglement.

\vspace*{0.3cm}
\noindent [2.5 points] c) Compute  the matrix that corresponds to the complete quantum circuit \\

consider $
1 \otimes \bf{H}  \equiv \begin{pmatrix}
\bf{H} &  \hat{\bf{0}} \\
\hat{\bf{0}} & {\bf{H}}
\end{pmatrix}
$. 

Confirm the results previously obtained when the input \acp{qubit} assumes the value $\vert00\rangle$.


%



\section*{Acknowledgment}

This work was partially supported by FCT (Fundação para a Ciência e a Tecnologia, Portugal), through the UIDB/50021/2020 project. The work reported in this paper has been performed by the author as professor responsible for the course {\it Technologies of Computing Systems} of the MSc program in Electrical and Computer Engineering at Instituto Superior Técnico, the faculty of Engineering of the University of Lisbon. 



\bibliographystyle{IEEEtran}
\bibliography{IEEEabrv,paper}
\end{document}